\input mtexsis
\paper
\singlespaced
\widenspacing
\twelvepoint
\Eurostyletrue
\thicksize=0pt
\sectionminspace=0.1\vsize
 
\def\discrete{\bf Z_2}

\def\yo1{{F_\pi^2}}

\def\sss{\scriptscriptstyle}

\def\dl{{ {\cal P}_{\sss L} }}
\def\dr{{ {\cal P}_{\sss R} }}

\def\bdr{{ {\bar {\cal P}}_{\sss R} }}

\def\ql{{ {\cal Q}_{\sss L} }}
\def\qr{{ {\cal Q}_{\sss R} }}
\def\pl{{ {\tilde {\cal Q}}_{\sss L} }}
\def\pr{{ {\tilde {\cal Q}}_{\sss R} }}
\def\bql{{ {\bar {\cal Q}}_{\sss L} }}

\def\bpl{{ {\bar {\tilde {\cal Q}}}_{\sss L} }}

\def\psil{{ {\cal Q}_{\sss L} }}
\def\psir{{ {\cal Q}_{\sss R} }}
\def\tpsil{{ {\cal P}_{\sss L} }}
\def\tpsir{{ {\cal P}_{\sss R} }}
\def\barpsil{{ {\bar Q}_{\sss L} }}
\def\barpsir{{ {\bar Q}_{\sss R} }}

\def\bartpsir{{ {\bar {\cal P}}_{\sss R} }}

\referencelist
\reference{ccwz} C.G.~Callan, S.~Coleman, J.~Wess and B.~Zumino,
              \journal Phys. Rev.;177,2247 (1969)
\endreference
\reference{witty}  E.~Witten,
              \journal Nucl. Phys. B;223,422 (1983)
\endreference
\reference{wz} J.~Wess and B.~Zumino,
              \journal Phys. Lett. B;37,95 (1971)
\endreference
\reference{hoker} E.~D'Hoker and S.~Weinberg, \journal Phys. Rev. D;50,6050 (1994), {\tt hep-ph/9409402}
\endreference
\reference{thooft}  G.~'t Hooft,
                    in {\it Recent Developments in Gauge Theories},
                    Proc. of the NATO ASI, Carg{\`e}se 1979,
                    ed. G.'t Hooft {\it et al} (Plenum, 1980)
\endreference
\reference{vafa}  C.~Vafa and E.~Witten,
              \journal Nucl. Phys. B;234,173 (1984)
\endreference
\reference{mano} A.~Manohar and G.~Moore, \journal Nucl. Phys. B;243,55 (1984)
\endreference
\reference{stein} J.~Steinberger, \journal Phys. Rev.;76,1180 (1949)
\endreference
\reference{detar}  C.~deTar and T.~Kunihiro,
\journal Phys. Rev. D;39,2805 (1989); 
Y.~Nemoto {\it et al}, {\tt hep-ph/9710445}
\endreference
\reference{itz} C.~Itzykson and J-B.~Zuber, {\it Quantum Field Theory} (McGraw-Hill, 1980)
\endreference
\reference{mended}  S.~Weinberg, \journal Phys. Rev. Lett.;65,1177 (1990);
           {\it ibid}, 1181
\endreference
\reference{beane} S.R.~Beane, \journal Ann. Phys.; 263,214 (1998), {{\tt hep-ph/9706246}}
\endreference
\reference{alg}  S.~Weinberg, \journal Phys. Rev.;177,2604 (1969)
\endreference
\endreferencelist
\titlepage
\obeylines
\hskip4.8in{DOE/ER/40762-153}
\hskip4.8in{U.ofMd.PP\#99-002}
\hskip4.8in{hep-th/9807116}
\unobeylines
\vskip0.5in
\title
A Note on Wess-Zumino Terms 
and Discrete Symmetries
\endtitle
\author
Silas R.~Beane

Department of Physics, University of Maryland
College Park, MD 20742-4111
\vskip0.1in
\center{{\it sbeane@pion.umd.edu}}\endcenter
\endauthor

\abstract
\singlespaced
\widenspacing
Sigma models in which the integer coefficient of the Wess-Zumino term
vanishes are easy to construct. This is the case if all flavor
symmetries are vectorlike. We show that there is a subset of
$SU(N)\times SU(N)$ vectorlike sigma models in which the Wess-Zumino
term vanishes for reasons of symmetry as well. However, there is no
chiral sigma model in which the Wess-Zumino term vanishes for reasons
of symmetry. This can be understood in the sigma model basis as a
consequence of an index theorem for the axialvector coupling
matrix. We prove this index theorem directly from the $SU(N)\times
SU(N)$ algebra.
\endabstract 
\vskip0.5in
\center{PACS: 11.30.Rd; 12.38.Aw; 12.90.+b; 11.30.Er} 
\endtitlepage
\vfill\eject                                     
\superrefsfalse
\singlespaced
\widenspacing

\vskip0.1in
\noindent {\twelvepoint{\bf 1.\quad Introduction}}
\vskip0.1in

Any microscopic theory which exhibits the pattern of spontaneous
symmetry breaking $SU(N)\times SU(N)\rightarrow SU(N)$ is described at
low energies by an effective theory of $N^2-1$ Goldstone bosons living
on the coset space $SU(N)\times SU(N)/SU(N)$.  Well known technology
tells how to build the most general lagrangian involving the Goldstone
bosons\ref{ccwz}. Introduce a field $U$ that transforms linearly with
respect to $SU(N)_L\times SU(N)_R$: $U\rightarrow {L} U {R^\dagger}$,
where ${{L,R}}$ is an element of $SU(N)_{L,R}$.  A convenient
parametrization of $U$ is

\offparens
$$
U=\exp{{2i{\pi_a}{T_a}}\over{F_\pi}}
\EQN def1
$$\autoparens 
where the $T_a$ are $SU(N)$ generators normalized such that
$Tr\,({T_a}{T_b})=\delta_{ab}/2$ and $\pi_a$ is the canonical
Goldstone boson field with decay constant $F_\pi$. In this basis the
Goldstone bosons transform nonlinearly with respect to $SU(N)_L\times
SU(N)_R$. Of the manifestly invariant terms in the lagrangian, the
unique term with the fewest derivatives is

\offparens
$$
{\textstyle{1\over 4}}{F_\pi^2}{\it Tr}\, 
{\partial_\mu}U{\partial^\mu}U^\dagger .
\EQN chirall2
$$\autoparens Consider the discrete symmetries of this
operator\ref{witty}. It is invariant with respect to charge
conjugation, $U\leftrightarrow{U^T}$, and parity, ${U_{\sss {\vec
x}}}\leftrightarrow{U_{\sss -{\vec x}}^\dagger}$. However, the
operator is separately invariant with respect to
${U}\leftrightarrow{U^\dagger}$, or $\pi\rightarrow -\pi$, which
counts modulo two the number of bosons.  Therefore this operator, as
well as all other manifestly invariant terms, has a discrete symmetry,
unrelated to $P$, $C$ and $T$, which forbids interactions of odd
numbers of Goldstone bosons. The absence of such a discrete symmetry
in realistic underlying theories like QCD can be used to motivate the
existence of other terms which involve interactions of odd numbers of
Goldstone modes\ref{witty}.

Although these so-called Wess-Zumino terms are not manifestly
invariant, they transform to a total derivative, which leaves the
action invariant\ref{wz}. They are the unique terms which transform to
a total derivative\ref{hoker}. General topological reasoning reveals
that the coefficient of such operators must be an
integer\ref{witty}. This argument follows from the topology of the
coset space and is therefore independent of details of the microscopic
theory. The contribution of the Wess-Zumino term to the action can be
written as

\offparens
$$
n{\Gamma_{\sss WZ}}
\EQN wzwno
$$\autoparens where $n$ is an integer\ref{witty}.  Note that
$\Gamma_{\sss WZ}$ is subject to usual operator rules of thumb. For
instance, it vanishes only if there is a symmetry ---such as
${U}\leftrightarrow{U^\dagger}$--- which forbids it. For this note it is
relevant to know that $\Gamma_{\sss WZ}$ contains interactions of odd
numbers of Goldstone bosons and the gauged Wess-Zumino term gives rise
to the process ${\pi_0}\rightarrow\gamma\gamma$.

The value of the integer $n$ is found by matching to the microscopic
theory. If $SU(N)\times SU(N)$ is chiral in the underlying gauge
theory then there are nonvanishing flavor anomalies. Here all axial
symmetries are by assumption spontaneously broken and therefore there
is no 't Hooft matching of anomalies between the microscopic theory
and the low-energy theory\ref{thooft}. However, anomalous Ward
identities must be satisfied.  The Wess-Zumino term fulfills this
role. If $SU(N)\times SU(N)$ is vectorlike in the underlying theory
then there are no flavor anomalies since a fermion with a mass term
allowed by a symmetry cannot contribute to an anomaly for that
symmetry. We then expect a vanishing integer in front of the
Wess-Zumino term. One might worry that the vectorlike scenario is in
contradiction with the Vafa-Witten theorem\ref{vafa}, which constrains
the breaking of vectorlike symmetries. This is not the case. A gauge
theory with an $SU(N)\times SU(N)$ vectorlike flavor symmetry does not
satisfy the Vafa-Witten positivity conditions. As we will see below,
there are either fundamental scalars or nonrenormalizable operators in
such a theory.

\figure{pizero}
\epsfxsize 4in
\centerline{\epsfbox{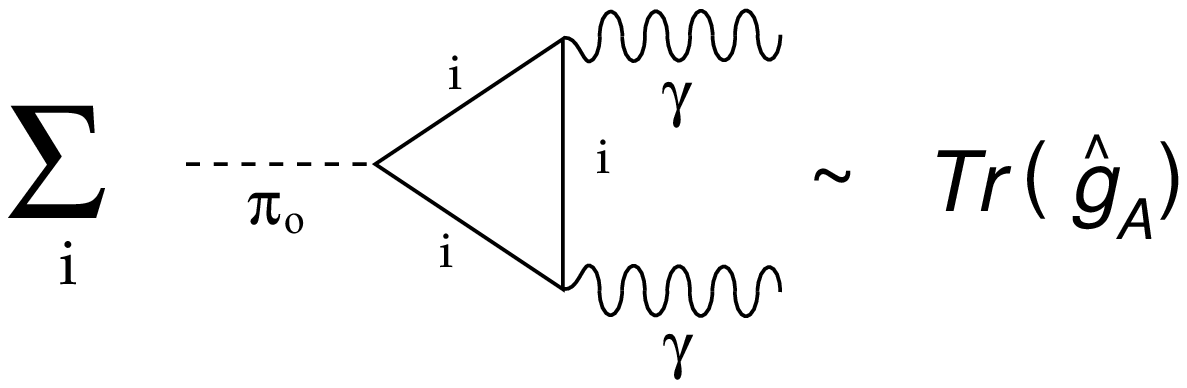}}
\caption{In a sigma model with arbitrary fermion content
the amplitude for ${\pi_0}\rightarrow\gamma\gamma$ is proportional to
$Tr\,({\hat g}_{\sss A})$.}
\endcaption
\endfigure

In QCD $SU(N)\times SU(N)$ is chiral and there are no discrete
symmetries beyond $P$, $C$ and $T$. Therefore, QCD is an example of an
underlying theory in which neither $n$ nor $\Gamma_{\sss WZ}$ vanish.
To our knowledge, other possibilities have not been considered in the
literature. In this note we construct models in which $n$ vanishes and
$\Gamma_{\sss WZ}$ does not vanish and in which $n$ and $\Gamma_{\sss
WZ}$ vanish. Not surprisingly the fermion content of these models is
vectorlike with respect to $SU(N)\times SU(N)$. We are unable to find
a more interesting model in which $n$ is nonzero and $\Gamma_{\sss
WZ}$ vanishes.  We will show that such a chiral model would be
inconsistent.

It is convenient to work with sigma models.  We focus entirely on
$SU(N)\times SU(N)$ sigma models with fermions in the fundamental
representation of $SU(N)$.  Given any sigma model, we can always
construct an asymptotically free gauge theory with precisely the same
symmetry structure and anomaly content. In a sigma model fermions
transform linearly with respect to $SU(N)\times SU(N)$ whereas
Wess-Zumino terms appear in the nonlinear realization.  Wess-Zumino
terms can be derived from the sigma models by taking into account the
effect of a change of basis on the path integral measure\ref{mano}.
The effect of the Wess-Zumino term in the sigma model basis can be
obtained by computing triangle graphs with fermions on the internal
lines. For instance, one can compute the amplitude for
${\pi_0}\rightarrow\gamma\gamma$\ref{stein}. Generally, in a sigma
model with Goldstone bosons and any number of fermions, this process
is proportional to $Tr\,({\hat g}_{\sss A})$ where ${\hat g}_{\sss A}$
is an axialvector coupling matrix (see \Fig{pizero}). In order to
reproduce the same physics in the two bases, it follows from
\Eq{wzwno} that

$$
Tr\,({\hat g}_{\sss A})=n 
\EQN indintro$$
where $n$ is the coefficient of the Wess-Zumino term. We will derive
this index theorem for ${\hat g}_{\sss A}$ directly from the
$SU(N)\times SU(N)$ algebra, and we will show that as a consequence of
this index theorem there is no chiral model with a discrete symmetry
which implies a vanishing Wess-Zumino term.

In section 2 we investigate several chiral sigma models; in
particular, we find the axialvector coupling matrices and search for
points of enhanced symmetry.  We investigate a vectorlike sigma model
in section 3. In section 4, we generalize our results and state an
index theorem for the axialvector coupling matrix. We show, as
consequence of the index theorem, that there is no chiral sigma model
with a discrete symmetry which implies a vanishing Wess-Zumino term.
We then prove the theorem and discuss other possible applications and
extensions. In section 5 we summarize.

\vskip0.1in
\noindent {\twelvepoint{\bf 2.\quad The Chiral Models}}
\vskip0.1in

\vskip0.1in
\noindent {\twelvepoint{\it 2.1\quad Model I}}
\vskip0.1in

The sigma models we will consider all contain a scalar field,
$\Sigma$, transforming in the $({\bar N},N)$ representation of
$SU(N)\times SU(N)$ together with various combinations of quark matter
transforming in the $(1,N)$ and $(N,1)$
representations$^1$\vfootnote1{Models similar to those discussed below
have been investigated in \Ref{detar}.}.  In this linear basis the
discrete symmetry we are searching for involves the transformation
$\Sigma\leftrightarrow{\Sigma^\dagger}$ together with a transformation
of the quark matter.  Consider first the ordinary sigma model. This
chiral model reflects the symmetries and anomaly structure of
$N$-flavor QCD. Its matter content consists of quarks and mesons
transforming as:

$$
\ruledtable
              | $SU(N)_L$  | $SU(N)_R$  \cr 
$\ql$       |  $\bf{N}$  | $\bf{1}$   \cr
$\qr$       |  $\bf{1}$   | $\bf{N}$  \cr
${\Sigma}$    | $\bf{\bar N}$  | $\bf{N}$  
\endruledtable
$$
with respect to $SU(N)\times SU(N)$. In our convention, $\ql={1\over
2}(1+{\gamma_5}){\cal Q}$ and $\qr={1\over 2}(1-{\gamma_5}){\cal Q}$.
The invariant interaction is

\offparens
$$
{\bql}{\Sigma}\qr +p.c.
\EQN chiral1$$\autoparens
It is clear that there is no additional discrete symmetry beyond QCD
symmetries. The axialvector coupling of this model is easy to find by
choosing $\Sigma ={F_\pi}U$ where $U$ is defined in \Eq{def1}.  This
choice allows us to ignore details of the symmetry breaking mechanism.
We find ${g}_{\sss A}=1$ and so ${\pi_0}\rightarrow\gamma\gamma$ is
nonvanishing and easily calculable from triangle graphs\ref{itz} for
any number of flavors. Therefore, the Wess-Zumino term computed from
this model by a change of basis is nonvanishing with $n=1$\ref{mano}.
The correct QCD result follows from assigning each quark a color
quantum number. In that case we have $n={N_c}{g}_{\sss A}={N_c}$.
Nonrenormalizable operators, such as

\offparens
$$
{\bql}{\Sigma}i{\slashchar{\partial}}{\Sigma^\dagger}\ql +p.c.
\EQN chiralsfd$$\autoparens
shift the value of ${g}_{\sss A}$. However, they do not contribute to
${\pi_0}\rightarrow\gamma\gamma$ in the chiral limit. Therefore they
have nothing to do with the Wess-Zumino term in the transformed basis.
It is sufficient to keep contributions to ${g}_{\sss A}$ from
renormalizable operators. The significance of the renormalizability
constraint will become clear in Section 4.

\vskip0.1in
\noindent {\twelvepoint{\it 2.2\quad Model II}}
\vskip0.1in

Adding more matter guarantees extra discrete symmetries for special
values of couplings. Hence, consider another chiral sigma model with
more matter, transforming as:

$$
\ruledtable
              | $SU(N)_L$  | $SU(N)_R$  \cr 
$\ql$, $\pl$       |  $\bf{N}$  | $\bf{1}$   \cr
$\qr$, $\pr$      |  $\bf{1}$   | $\bf{N}$  \cr
${\Sigma}$    | $\bf{\bar N}$  | $\bf{N}$  
\endruledtable
$$
with respect to $SU(N)\times SU(N)$.  The invariant interaction is

\offparens
$$
a({\bql}{\Sigma}\qr +p.c.)+
b({\bpl}{\Sigma}\pr +p.c.)+
c({\bql}{\Sigma}\pr + {\bpl}{\Sigma}\qr +p.c.).
\EQN chiral2$$\autoparens
With $a=b$ this model has a discrete symmetry with respect to
interchange of $\ql$ and $\pr$, $\qr$ and $\pl$, and $\Sigma$ and
$\Sigma^\dagger$. This symmetry is a generalization of parity and
leads to no new multiplicatively conserved quantum number.  In this
model we find,

$$
{{\hat g}_{\sss A}}=
\left(\matrix{1  & 0 \cr
              0  & 1     }\right)
\EQN chiral2ga$$
for all values of the couplings $a$, $b$ and $c$.  Therefore
$Tr\,({\hat g}_{\sss A})=2$, and ${\pi_0}\rightarrow\gamma\gamma$ is
nonvanishing and twice the value obtained in the ordinary sigma
model. Hence the Wess-Zumino term has coefficient $n=2$.

\vskip0.1in
\noindent {\twelvepoint{\it 2.3\quad Model III}}
\vskip0.1in

The chiral models considered above have no nontrivial mixing because
there are no invariant quark bilinears.  Consider therefore a chiral
sigma model with yet more matter, transforming as:

$$
\ruledtable
              | $SU(N)_L$  | $SU(N)_R$  \cr 
$\ql$, $\pl$, $\dr$       |  $\bf{N}$  | $\bf{1}$   \cr
$\qr$, $\pr$, $\dl$      |  $\bf{1}$   | $\bf{N}$  \cr
${\Sigma}$    | $\bf{\bar N}$  | $\bf{N}$  
\endruledtable
$$
with respect to $SU(N)\times SU(N)$. The invariant interaction is:

\offparens
$$\eqalign{&
a({\bql}{\Sigma}\qr +p.c.)+
b({\bpl}{\Sigma}\pr +p.c.)+
c({\bql}{\Sigma}\pr + {\bpl}{\Sigma}\qr +p.c.)\cr
&d({\bdr}{\Sigma}\dl +p.c.)+
{M_0}({\bdr}{\ql}+ {\bql}{\dr}+p.c.)+
{{\tilde M}_0}({\bdr}{\pl}+ {\bpl}{\dr}+p.c.).\cr}
\EQN chiral2a$$\autoparens
Note the invariant bilinears. In the symmetric phase, two of the
quarks are massive and degenerate and one is massless. Here there is a
discrete symmetry with respect to interchange of $Q$ and $P$ and
$\Sigma$ and $\Sigma^\dagger$ when $a=d$, $b=c={{\tilde M}_0}=0$, and
with respect to interchange of $\tilde Q$ and $P$ and $\Sigma$ and
$\Sigma^\dagger$ when $b=d$, $a=c={M_0}=0$. However these choices of
parameters correspond to decoupling $\tilde Q$ and $Q$, respectively,
which yields the simplest vectorlike model for the remaining
matter. We will study this model in the next section.  In this model
we find

$$
{{\hat g}_{\sss A}}=
\left(\matrix{ \cos^2\gamma\cos 2\beta +\sin^2\gamma & 
     \cos\gamma\sin\gamma(\cos 2\beta -1) & \cos\gamma\sin 2\beta \cr
     \cos\gamma\sin\gamma(\cos 2\beta -1) &  
   \sin^2\gamma{\cos 2\beta}+\cos^2\gamma
     & -\sin\gamma\sin 2\beta     \cr
    \cos\gamma\sin 2\beta   & -\sin\gamma\sin 2\beta  
     & -\cos 2\beta  }\right),
\EQN chiral3ga$$
where the mixing angles are related to the coupling parameters.  Note
that $({\hat g}_{\sss A})^2={\bf 1}$. This is a statement of the
Adler-Weisberger sum rule.  If for special values of the couplings
there is a discrete symmetry which implies a multiplicatively
conserved quantum number and involves $\pi\rightarrow -\pi$, then the
diagonal elements of ${\hat g}_{\sss A}$ must vanish for the
corresponding values of the mixing angles. But note that $Tr\,({\hat
g}_{\sss A})=1$, independent of the values of the mixing angles. This
is of course consistent with the index theorem.  This model therefore
has no relevant discrete symmetry, as deduced directly from the
lagrangian. The amplitude for ${\pi_0}\rightarrow\gamma\gamma$ is the
same as in the ordinary sigma model (Model I).  We will see below that
generally there is no chiral model with the extra discrete
symmetry. But first we will consider the simplest vectorlike model.

\vskip0.1in
\noindent {\twelvepoint{\bf 3.\quad The Vectorlike Model}}
\vskip0.1in

The matter content of the vectorlike model consists of quarks and
mesons transforming as:

$$
\ruledtable
              | $SU(N)_L$  | $SU(N)_R$  \cr 
$\psil$, $\tpsir$  |  $\bf{N}$  | $\bf{1}$    \cr
$\psir$, $\tpsil$  |  $\bf{1}$   | $\bf{N}$   \cr
${\Sigma}$           | $\bf{\bar N}$   | $\bf{N}$  
\endruledtable
$$
with respect to $SU(N)\times SU(N)$. Note that there are equal numbers
of left- and right-handed Weyl fermions assigned to each charge in a
vectorlike model. This guarantees that all quarks have invariant
masses. The invariant interaction is

\offparens
$$
a({\barpsil}{\Sigma}\psir +p.c.)+
b({\bartpsir}{\Sigma}\tpsil +p.c.)+
{M_0}({\barpsil}\tpsir + {\barpsir}\tpsil +p.c.).
\EQN vector1$$\autoparens
All fermions are massive.  The free fermion theory is $U(2N)\times
U(2N)$ invariant.  The invariant mass term breaks this to $U(2N)$, and
the Yukawa interactions further break this to $SU(N)\times SU(N)\times
U(1)$.  Therefore, unlike the ordinary chiral sigma model where a
microscopic theory in which the fermions have only gauge interactions
can always be constructed, here a microscopic theory must have
additional nongauge interactions. It must therefore have fundamental
scalars or nonrenormalizable interactions. In neither case does the
Vafa-Witten theorem constrain the pattern of symmetry
breaking\ref{vafa}.  Vectorlike $SU(N)\times SU(N)$ sigma models are
therefore worthy of study.

Here we have an additional discrete symmetry if $a=b$.  If $a=b$ there
is a $\discrete$ symmetry corresponding to interchange of ${\cal Q}$
and ${\cal P}$ and $\Sigma$ and $\Sigma^\dagger$. This symmetry
commutes with parity and so we can assign multiplicatively conserved
charges to each physical state:

$$
\ruledtable
              | P  | $\discrete$  \cr 
$\psi_+$         | $\;\;\, 1$  |      $\;\;\, 1$      \cr
$\psi_-$         | $\;\;\, 1$  |     $-1$      \cr
$\pi$         | $-1$  |     $-1$      
\endruledtable
$$
Here $\psi_\pm$ denote the quarks in the diagonal basis.  This is
precisely the discrete symmetry which rules out self interactions of
odd numbers of Goldstone bosons.  In this model we find

$$
{{\hat g}_{\sss A}}=
\left(\matrix{-\cos{2\phi}  & -\sin{2\phi} \cr
    -\sin{2\phi}  & \quad\cos{2\phi}     }\right),
\EQN vecga1$$
where $\cot 2\phi =(a-b){F_\pi}/2{M_0}$. Again $({\hat g}_{\sss
A})^2={\bf 1}$, a statement of the Adler-Weisberger sum rule. Since
$Tr\,({\hat g}_{\sss A})=0$, ${\pi_0}\rightarrow\gamma\gamma$ vanishes
for all values of $a$ and $b$, as expected in a vectorlike model.  In
the transformation to the nonlinear basis, the nontrivial Jacobians
due to ${\cal Q}$ and ${\cal P}$ cancel.  Note that if $a=b$,

$$
{{\hat g}_{\sss A}}=
\left(\matrix{0  & 1 \cr
       1  & 0     }\right).
\EQN vecga2$$
The diagonal elements of the axialvector coupling matrix vanish
because they do not respect the $\discrete$ symmetry. That is,
${g_{\sss \pi{\psi_-}{\psi_-}}}={g_{\sss \pi{\psi_+}{\psi_+}}}=0$.  The
Adler-Weisberger sum rule then predicts the off-diagonal axialvector
couplings.  At this point of enhanced symmetry, all triangle graphs
vanish (and not just the sum). Correspondingly, the Wess-Zumino
operator in the nonlinear basis vanishes (and not just the integer
coefficient). Clearly this is not of great practical interest.

This sigma model provides a field theoretic realization of Weinberg's
theory of the axialvector coupling of the constituent
quark\ref{mended}. The vectorlike nature of the model and the point of
enhanced $\discrete$ symmetry are in precise correspondence with
algebraic sum rules for Goldstone boson scattering, which are well
satisfied in nature\ref{beane}. This correspondence is a profound
puzzle.

\vskip0.1in
\noindent {\twelvepoint{\bf 4.\quad An Index Theorem}}
\vskip0.1in

The results of the sigma models considered above can be summarized by
the index theorem:

\offparens
$$
Tr\,({\hat g}_{\sss A})= {n_{\sss L}}-{n_{\sss R}}
\EQN indthm$$
where, by convention, ${n_{\sss {L}}},{n_{\sss {R}}}$ is the number of
left-handed Weyl fermions with $SU(N)\times SU(N)$ charge
$(N,1),(1,N)$.  In a chiral sigma model ${n_{\sss L}}\neq{n_{\sss R}}$
and in a vectorlike model ${n_{\sss L}}={n_{\sss R}}$, by definition.
If $\pi\rightarrow -\pi$ with respect to a discrete symmetry which
implies a multiplicatively conserved charge, then the diagonal
elements of ${\hat g}_{\sss A}$ transform like $\pi$.  Therefore in a
model with such a symmetry, $[{\hat g}_{\sss A}]_{\sss\alpha\alpha}=0$
$\forall$ $\alpha$, and clearly $Tr\,({\hat g}_{\sss A})= 0$.  Such a
model is vectorlike. Therefore, there can be no chiral sigma model
with such a discrete symmetry.

It is important to know precisely what enters \Eq{indthm}.  Here we
give a proof of \Eq{indthm} that is purely a consequence of the
$SU(N)\times SU(N)$ representation theory. We emphasize that the index
theorem in the linear basis has nothing to do with topology.  Assume
that we are in a helicity conserving Lorentz frame and that the
quarks, for each helicity $\lambda$, fall into a reducible
representation of $SU(N)\times SU(N)$ consisting of a sum of any
number of $(1,N)$ and $(N,1)$ representations. For a given helicity,
we can express the diagonal quark states as

$$
\ket{\psi}^\lambda_{\sss\alpha}= 
{\sum_{j=1}^{n_{\sss L}}} U_{{\sss\alpha}j}{\ket{N,1}_j} +
{\sum_{{\bar k}=1}^{n_{\sss R}}} V_{{\sss\alpha}{\bar k}}{\ket{1,N}_{\bar k}} 
\EQN bdef$$
where $\hat U$ and $\hat V$ are mixing matrices and $\alpha$ ranges
from $1$ to ${n_{\sss L}}+{n_{\sss R}}$. The parity operator in the
helicity conserving frame acts as $P\ket{\psi}^\lambda_{\sss\alpha}=
\ket{\psi}^{-\lambda}_{\sss\alpha}$ and as 
$P{\ket{\alpha ,\beta}_{i}}={\ket{\beta,\alpha}_{i}}$ and
$P{\ket{\alpha ,\beta}_{\bar k}}={\ket{\beta,\alpha}_{\bar k}}$ on
states of definite $SU(N)\times SU(N)$.  The $SU(N)\times SU(N)$
algebra can be expressed as

$$
[{{\cal Q}^5_a},{{\cal Q}^5_b}]=i{f_{abc}}{T_c}\qquad
[{T_a},{{\cal Q}^5_b}]=i{f_{abc}}{{\cal Q}^5_c}\qquad
[{T_a},{T_b}]=i{f_{abc}}{T_c}
\EQN axchdef$$
where ${\cal Q}^5_b$ is the axialvector charge operator and the
${T_c}$ are $SU(N)$ generators. The action of the axialvector charge
operator on the states of definite $SU(N)\times SU(N)$ is

$$
{{\cal Q}^5_a}\ket{1,N}=-{T_a}\ket{1,N}\qquad
{{\cal Q}^5_a}\ket{N,1}={T_a}\ket{N,1}.
\EQN qdef$$
This sign convention is in agreement with \Eq{indthm}.  The quark
axialvector coupling matrix is then defined by

$$
{}^\lambda_{\sss\alpha}\bra{\psi}{{\cal Q}^5_a}\ket{\psi}^\lambda_{\sss\beta}=
[{\hat g}^\lambda_{\sss A}]_{\sss\alpha\beta}{T_a}.
\EQN gadef$$
The results of the models of the previous sections for ${\hat g}_{\sss
A}$ are easily recovered by choosing values of ${n_{\sss L}}$ and
${n_{\sss R}}$ and a representation of the $\hat U$ and $\hat V$
mixing matrices. For instance, model I corresponds to ${n_{\sss
L}}=1$, ${n_{\sss R}}=0$ and $U=1$. Model III corresponds to ${n_{\sss
L}}=2$, ${n_{\sss R}}=1$, and $\hat U$ and $\hat V$ a combination of
Euler angles. The vectorlike model corresponds to ${n_{\sss
L}}={n_{\sss R}}=1$, ${U_1}={V_2}=\sin\phi$ and
${V_1}=-{U_2}=\cos\phi$.  Therefore, all that is of interest in the
sigma models is a consequence of the $SU(N)\times SU(N)$
representation theory.  We are now in a position to prove the index
theorem.

\noindent {\underbar {\it Proof:}}

\noindent Orthonormality of the $\ket{\psi}_{\sss\alpha}$ and the 
definition of ${\hat g}_{\sss A}$ give, respectively:

\offparens
$$\EQNalign{
&{\sum_{j=1}^{n_{\sss L}}} U^*_{{\sss\alpha}j}U_{{\sss\beta}j}+
{\sum_{{\bar k}=1}^
{n_{\sss R}}} V^*_{{\sss\alpha}{\bar k}}V_{{\sss\beta}{\bar k}}=
\delta_{\sss\alpha\beta}\EQN basic;a \cr 
&{\sum_{j=1}^{n_{\sss L}}} U^*_{{\sss\alpha}j}U_{{\sss\beta}j}-
{\sum_{{\bar k}=1}^{n_{\sss R}}} 
V^*_{{\sss\alpha}{\bar k}}V_{{\sss\beta}{\bar k}}=
[{\hat g}^\lambda_{\sss A}]_{{\sss\alpha}{\sss\beta}}.
\EQN basic;b \cr} 
$$\autoparens It is easy to check that the generalized
Adler-Weisberger sum rule, $[{\hat g}^\lambda_{\sss
A}]_{{\sss\alpha}{\sss\beta}} [{\hat g}^\lambda_{\sss
A}]_{{\sss\beta}{\sss\gamma}}= \delta_{\sss\alpha\gamma}$, is
satisfied.  Since $Tr\,({\delta_{{\sss\alpha}{\sss\beta}}})= {n_{\sss
L}}+{n_{\sss R}}$,
\Eq{basic;a} and the independence of left and right give:

$$
Tr\,{\sum_{j=1}^{n_{\sss L}}} U^*_{{\sss\alpha}j}U_{{\sss\beta}j}=
{n_{\sss L}}\qquad
Tr\,{\sum_{{\bar k}=1}^{n_{\sss R}}} 
V^*_{{\sss\alpha}{\bar k}}V_{{\sss\beta}{\bar k}}={n_{\sss R}}.
\EQN whatever$$
It then follows directly from \Eq{basic;b} that

\offparens
$$
Tr\,({\hat g}^{\pm\lambda}_{\sss A})= \pm ({n_{\sss L}}-{n_{\sss R}})
\EQN indthm2$$\autoparens
with $+\lambda$ identified with left-handed helicity and $-\lambda$
identified with right-handed helicity, which is the desired
relation. {\it Q.E.D.}

It might seem puzzling that the index theorem follows purely from
$SU(N)\times SU(N)$ while the sigma model results rely on
renormalizability. There is no paradox; the full $SU(N)\times SU(N)$
symmetry in helicity conserving frames implies a constraint on the
asymptotic behaviour of Goldstone boson scattering amplitudes in the
broken phase\ref{alg}.  Evidently renormalizability, which is of
course also a statement about high energy behavior, yields an
equivalent constraint in this context\ref{mended}.

Does the index theorem generalize to higher representations of
$SU(N)\times SU(N)$?  For instance, in $N=2$ QCD, baryons fall into
reducible representations which are sums of any number of $(1,2)$,
$(2,1)$ {\it and} $(1,4)$, $(4,1)$, $(3,2)$ and $(2,3)$
representations.  The four-dimensional representations contain the
isospin $3/2$ baryons.  It is not clear to the author whether an
extension of the index theorem to include these representations is
possible or even sensible.

\vskip0.1in
\noindent {\twelvepoint{\bf 5.\quad Summary}}
\vskip0.1in

In the nonlinear realization of $SU(N)\times SU(N)/SU(N)$, manifestly
invariant operators in the lagrangian have a discrete symmetry which
rules out interactions of odd numbers of Goldstone
bosons\ref{witty}. The fact that this discrete symmetry is not a
symmetry of realistic theories like QCD motivates the existence of
additional operators which break this discrete symmetry\ref{wz}.
These Wess-Zumino operators are not manifestly invariant and yet they
transform to a total derivative and therefore leave the action
invariant. The coefficient of these operators is an integer, which is
determined in practice by matching to the underlying theory.
Generally in underlying theories like QCD in which $SU(N)\times SU(N)$
is chiral, this integer is nonvanishing. In this letter we have
searched for a chiral model in which the integer of the Wess-Zumino
term is nonvanishing and yet has a discrete symmetry, $\pi\rightarrow
-\pi$, which implies that the Wess-Zumino term must vanish.

We found it convenient to work with sigma models in which the fermion
fields transform linearly with respect to $SU(N)\times
SU(N)$. Wess-Zumino terms can be derived in these models by a
transformation to a nonlinear basis\ref{mano}. In the linear models
the effects of anomalies are contained in triangle graphs\ref{stein}.
The fact that the coefficient of the Wess-Zumino term is an integer
has an analogous statement in the linear basis. In the linear basis,
the trace of the axialvector coupling matrix, which determines the sum
of all triangle graphs, must be an integer. We proved this index
theorem directly from the $SU(N)\times SU(N)$ algebra.  In chiral
models the integer is nonvanishing and in vectorlike models the
integer vanishes.  If a given sigma model has a discrete symmetry
which implies a multiplicatively conserved charge and if
$\pi\rightarrow -\pi$ with respect to this discrete symmetry, then
each diagonal element of the axialvector coupling matrix must
vanish. In such a case the trace vanishes. Therefore a model with this
discrete symmetry must be vectorlike with respect to $SU(N)\times
SU(N)$. We conclude that there is no chiral model with a discrete
symmetry which implies a vanishing Wess-Zumino term.

\vskip0.1in

\noindent This work was supported by the U.S. Department of Energy grant
DE-FG02-93ER-40762. I thank Markus Luty for valuable conversations.

\nosechead{References}
\ListReferences \vfill\supereject \end